\documentclass[reprint,superscriptaddress,amsmath,amssymb,aps,prresearch]{revtex4-2}
\usepackage{graphicx}
\usepackage{dcolumn}
\usepackage{bm}
\usepackage{braket}
\usepackage{booktabs}
\usepackage[table]{xcolor}
\usepackage{colortbl}
\usepackage{etoolbox}
\usepackage{placeins}
\usepackage[abs]{overpic}
\makeatletter
\newlength{\qrr@dimen@}
\expandafter\pretocmd\csname tabular*\endcsname{\setlength{\qrr@dimen@}{#1}}{}{}
\newcommand*{\Rowcolor}[2][\tabcolsep]{%
    \ifx\relax#1\relax\else
        \kern-\the\dimexpr#1\relax
    \fi
    \makebox[0pt][l]{%
        \fboxsep=0pt
        \colorbox{#2}{%
            \strut\kern\qrr@dimen@
        }%
    }%
    \ifx\relax#1\relax\else
        \kern\the\dimexpr#1\relax
    \fi
    \ignorespaces
}
\usepackage[colorlinks,urlcolor=blue,citecolor=blue,linkcolor=blue]{hyperref}

\begin{document}

\title{O-Sensing: Operator Sensing for Interaction Geometry and Symmetries}

\author{Ye-Ming Meng}
\affiliation{State Key Laboratory of Precision Spectroscopy, East China Normal University, Shanghai 200062, China}
\author{Zhe-Yu Shi}
\email{zyshi@lps.ecnu.edu.cn}
\affiliation{Institute of Quantum Science and Precision Measurement, School of Physics, East China Normal University, Shanghai 200062, China}

% \date{\today}

\begin{abstract}
    We ask whether the Hamiltonian, interaction geometry, and symmetries of a quantum many-body system can be inferred from a few low-lying eigenstates without knowing which sites interact with each other.
    Directly solving the eigenvalue equations imposes constraints that yield a highly degenerate subspace of candidate operators, where the local Hamiltonian is hidden among an extensive family of conserved quantities, obscuring the interaction geometry.
    Here we introduce O-Sensing, a protocol designed to extract the Hamiltonian and symmetries directly from these states.
    Specifically, O-Sensing employs parsimony-driven optimization to extract a maximally sparse operator basis from the degenerate subspace.
    The Hamiltonian is then selected from this basis by maximizing spectral entropy (effectively minimizing degeneracy) within the sampled subspace.
    We validate O-Sensing on Heisenberg models on connected Erd\H{o}s--R\'enyi graphs,
    where it reconstructs the interaction geometry and uncovers additional long-range conserved operators.
    We establish a learnability phase diagram across graph densities, featuring a pronounced ``confusion'' regime where parsimony favors a dual description on the complement graph.
    These results show that sparsity optimization can reconstruct interaction geometry as an emergent output,
    enabling simultaneous recovery of the Hamiltonian and its symmetries from low-energy eigenstates.
\end{abstract}

\maketitle

\section{Introduction}
The principle of parsimony—the idea that nature's laws are distinguished by maximal simplicity—has long underpinned the pursuit of physical discovery~\cite{Einstein1934, Schmidt2009}.
In most standard frameworks, spatial geometry serves as a fixed background, a pre-existing stage on which physical degrees of freedom reside.
By contrast, a modern viewpoint treats geometry not as an a priori background~\cite{Zanardi2004, Ryu2006, Raasakka2017, Pollack2018},
but as an emergent structure encoded in the correlation patterns of quantum states~\cite{Ryu2006, van2010, Eisert2010, Evenbly2011, Swingle2012, Cao2017, Carroll2022}.
In this view, the ``space'' we perceive is the coordinate system in which the governing laws take their simplest and most local form~\cite{Zanardi2004, Qi2013, Cotler2019, Carroll2021, Loizeau2023, Soulas2025}.
Within the emerging framework of data-driven quantum discovery~\cite{Carleo2019, Carrasquilla2020, Huang2022},
one can ask whether the Hamiltonian, interaction geometry, and hidden symmetries can be reconstructed solely from observations.

This question is closely related to the broader program of Hamiltonian learning: reconstructing a parent Hamiltonian from stationary states, thermal ensembles, or nonequilibrium dynamics~\cite{Chertkov2018, Garrison2018, Qi2019, Rattacaso2024, Bairey2019, Huang2023}.
A representative method is the Eigenstate-to-Hamiltonian Construction (EHC)~\cite{Chertkov2018} (also independently formulated in Ref.~\cite{Qi2019}), which turns the eigenvalue equation into linear constraints, yielding a kernel subspace of operators consistent with the data and containing the parent Hamiltonian.
Crucially, however, these constraints are satisfied by any conserved quantity. The resulting kernel therefore captures not just the parent Hamiltonian, but the entire algebra of symmetries and their mixtures, forming a highly degenerate subspace of valid solutions.
Furthermore, when the interaction geometry is unknown, the lack of spatial priors forces one to search within an all-to-all operator basis~\cite{Chertkov2018}, dramatically expanding the dimensionality of the candidate manifold.

Navigating this subspace is challenging because a generic element of the kernel is a dense algebraic mixture of the Hamiltonian and symmetry operators.
While such mixtures remain consistent with the observations, they typically obscure local structure and lack physical interpretability.
Recovering the parent Hamiltonian therefore requires selecting a distinguished basis within this degenerate subspace that minimizes complexity.
This pursuit of parsimony parallels classical system identification~\cite{Brunton2016, Lusch2018} and statistical learning~\cite{Tibshirani1996, Candes2006}, where structural sparsity is key to extracting interpretable models.
While sparsity-driven inference is well-established in classical systems, its application to non-commutative operator algebras is far less developed, and the present work offers a concrete procedure in this setting.

\begin{figure*}
    \centering
    \begin{overpic}[width=.94\linewidth]{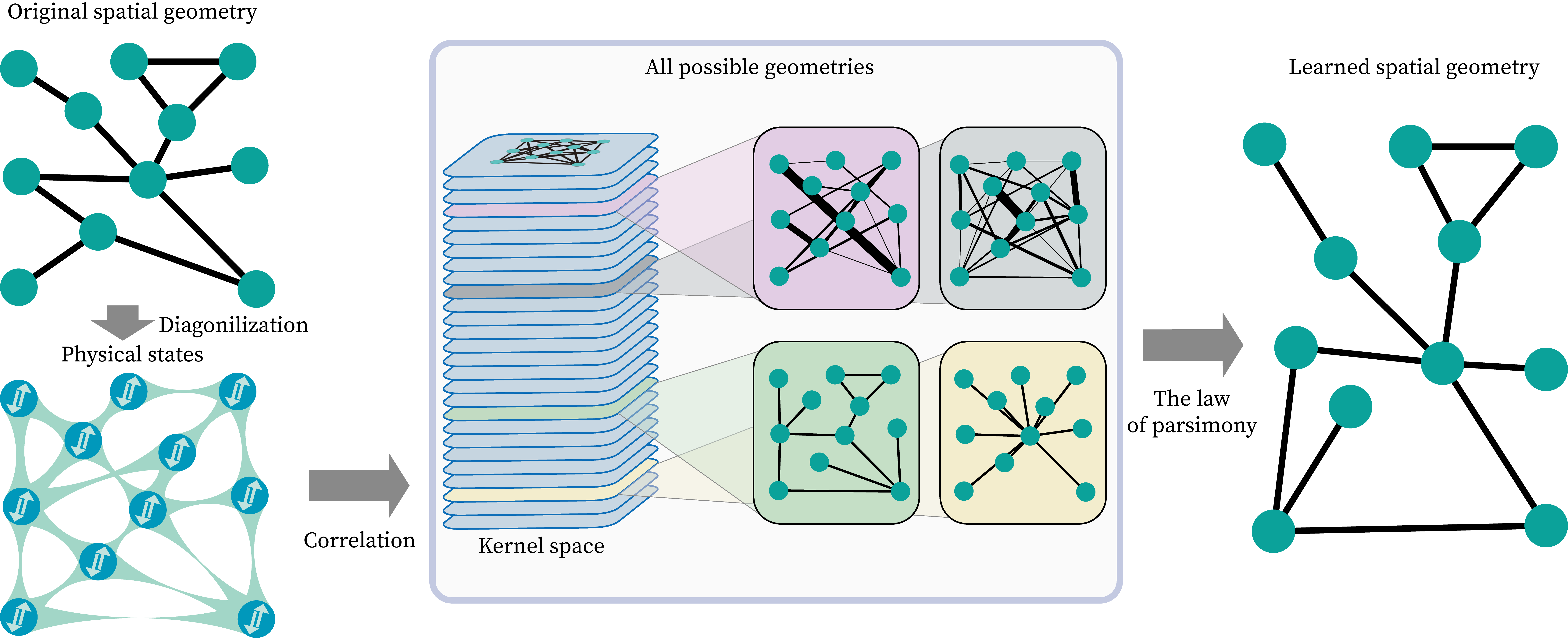}
        \put(-18,185){\textbf{(a)}}
        \put(-18,75){\textbf{(b)}}
        \put(100,185){\textbf{(c)}}
        \put(350,185){\textbf{(d)}}
    \end{overpic}
    \caption{Learning spatial geometry and symmetries from low-energy states of a quantum system.
    (a) The original spatial geometry, depicted as a random, connected Erd\H{o}s--R\'enyi graph. Circles denote physical sites, while lines represent interactions.
    (b) Solving the eigenvalue problem yields low-energy states, revealing correlations between sites. Directly reading the original geometry from these correlations is infeasible since two spatially separated sites may be correlated.
    (c) The parent operator subspace derives a linear space of operators satisfying the eigenvalue equation. Randomly chosen basis (top half) obscure physical information, making interpretation difficult. Optimizing the basis (bottom half) identifies the sparsest basis, which optimally describes the system with clear physical meaning. This optimized basis includes the parent Hamiltonian, encoding the spatial geometry, and conserved quantities representing system symmetries.
    (d) Spatial geometry learned via sparse basis optimization of the parent operator subspace. The method reconstructs the original Hamiltonian with high precision, successfully retrieving the geometry.
    } \label{fig1}
\end{figure*}
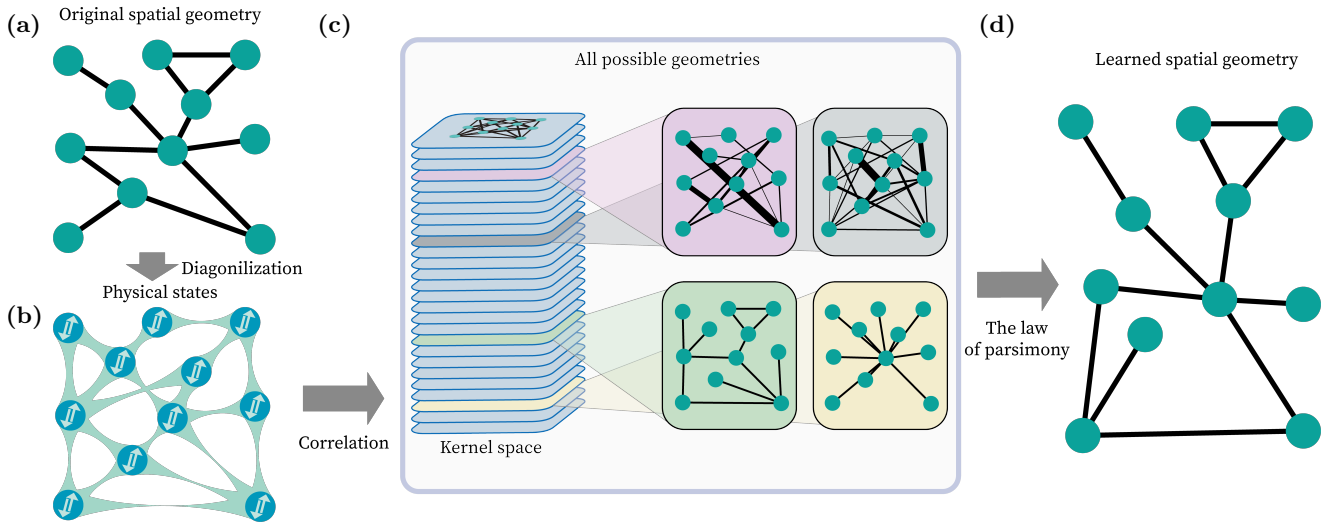
% \FloatBarrier

In this work, we formalize the principle of parsimony into a practical protocol termed \textit{O-Sensing} (Operator Sensing).
Specifically, O-Sensing resolves the degeneracy of the EHC kernel---which we call the \emph{parent operator subspace}---through a two-step protocol.
First, to disentangle the dense operator mixtures, we draw inspiration from sparse optimization in statistical learning~\cite{Tibshirani1996, Chen2001, Zou2006} and signal processing~\cite{Candes2006, Donoho2006, Zhang2015}.
We show that, when operators are represented in a local basis, parsimony becomes operational as sparsity: the physically interpretable generators are those with the sparsest coefficient representations. Treating sparsity as a ``gauge-fixing'' principle, O-Sensing transforms the parent subspace into a distinguished basis, allowing the interaction geometry to naturally emerge as the coordinate system that minimizes complexity.
Second, to pinpoint the parent Hamiltonian among these recovered sparse generators, we introduce a spectral-entropy criterion.
Because symmetries induce high spectral degeneracy, they minimize spectral entropy; the Hamiltonian, in contrast, is uniquely distinguished by maximizing spectral resolution over the Hilbert space within the sampled low-energy subspace.

We test O-Sensing on Heisenberg models on connected Erd\H{o}s--R\'enyi graphs, where the interaction geometry is known but not provided to the algorithm.
From low-energy correlations alone, our approach reconstructs the interaction connectivity, identifies both intrinsic and geometric symmetries, and uncovers additional hidden long-range conserved quantities.
Our results demonstrate that sparsity optimization serves as an effective tool for extracting interaction structures from non-local quantum observables, even without prior knowledge of the underlying geometry.

\section{The O-Sensing Framework}
\subsection{Hidden Geometry on Interaction Graphs}
We operationalize the recovery of emergent geometry by considering a spin-$\frac{1}{2}$ system whose degrees of freedom reside on the vertices of a generic interaction graph (Fig.~\ref{fig1}a).
As a minimal setting in which locality is well-defined while the geometry is irregular, we focus on the Heisenberg Hamiltonian:

\begin{equation}\label{eq:hamiltonian}
    H = \sum_{\langle i,j\rangle \in \text{edges}} J_{ij} \bm{\sigma}_i \cdot \bm{\sigma}_j,
\end{equation}
where $\bm{\sigma}_i = (\sigma_i^x, \sigma_i^y, \sigma_i^z)$ are Pauli matrices at site $i$, and $J_{ij}<0$ ($>0$) denotes ferromagnetic (antiferromagnetic) couplings. For regular lattices (e.g., 1D chains and 2D square, triangular, or honeycomb lattices), locality is tied to a metric distance; here it is defined purely by edge adjacency on the interaction graph.

To test geometry recovery in the absence of spatial priors, we define an Erd\H{o}s--R\'enyi interaction graph $G(N_v,N_e)$~\cite{Erdos1959, Bollobas2001}, constructed by sampling $N_e$ edges uniformly at random from the $\binom{N_v}{2}$ possible vertex pairs and conditioning the resulting graph to be connected.
Treating the interaction geometry as unknown, our objective is to reconstruct the parent Hamiltonian and hidden symmetries solely from low-energy observables, thereby recovering the underlying connectivity.

The reconstruction protocol begins with the computation of correlation matrices~\cite{Qi2019} from a set of low-energy eigenstates $\{\ket{\Psi_\alpha}\}$.
Although the governing Hamiltonian is local, the resulting states typically exhibit long-range correlations that extend far beyond the interaction range (Fig.~\ref{fig1}b).
Consequently, the interaction geometry cannot be directly read out from the correlation data; instead, it must be retrieved by reconstructing interaction terms of the parent Hamiltonian.

\subsection{Parent Operator Subspace}
\begin{figure}
    \includegraphics[width=.98\linewidth]{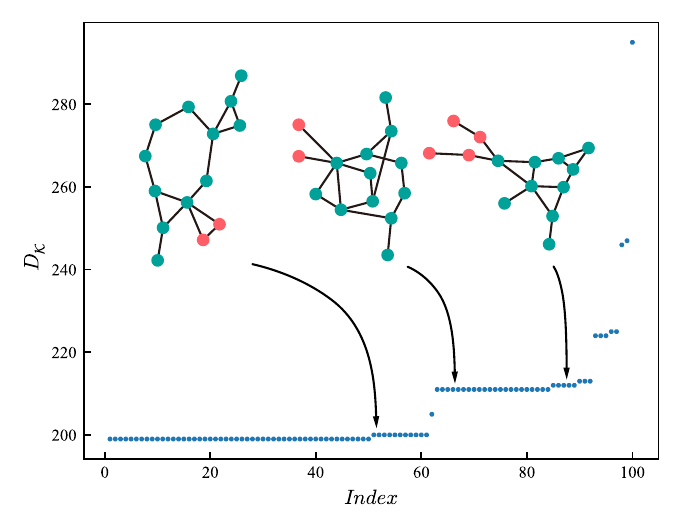}
    \caption{Parent operator subspace dimension $D_{\mathcal{K}}$ for an ensemble of 100 connected Erd\H{o}s--R\'enyi graphs ($N_v=14, N_e=17$) with uniform antiferromagnetic coupling. Individual samples are ordered along the horizontal axis by their respective kernel dimensions. The results exhibit a hierarchy of discrete plateaus, starting from the universal baseline of $D_{\mathcal{K}} = N_v^2 + 3 = 199$.}
    \label{fig2}
\end{figure}

To reconstruct the parent Hamiltonian and conserved quantities, we first define the linear space of operators consistent with the eigenvalue equation $O\ket{\Psi_\alpha}=\lambda_{\alpha}\ket{\Psi_\alpha}$ for a set of sampled eigenstates indexed by $\alpha$.
Since the interaction geometry is unknown, any candidate operator $O$ is expanded in a graph-independent operator basis $\{W_i\}$,
\begin{equation}
    O = \sum_{i=1}^{N_\mathcal{W}} c_i W_i .
\end{equation}
Here $\mathcal{W}$ consists of Hermitian few-body operators (up to three sites)~\footnote{We restrict to operators that conserve the total $\sigma^z$.
Concretely, we include $\sigma_i^z$, $\sigma_i^z\sigma_j^z$, and $\sigma_i^+\sigma_j^-+\sigma_j^+\sigma_i^-$ for all $i<j$, as well as $\sigma_i^z\sigma_j^z\sigma_k^z$ and $(\sigma_i^+\sigma_j^-+\sigma_j^+\sigma_i^-) \sigma_k^z$ for all distinct triplets. We define $\sigma^\pm \equiv (\sigma^x \pm i\sigma^y)/2$. For $N_v=14$, this yields $14$ one-body operators, $182$ two-body operators, and $1456$ three-body operators.} 
so that each $O$ is uniquely identified by its coefficient vector $\mathbf{c}=(c_1,\dots,c_{N_\mathcal{W}})^T$.
Since $\{W_i\}$ are chosen as Hermitian and we seek real coefficients $c_i$, the resulting operator $O$ is Hermitian by construction.
The eigenvalue equation $O\ket{\Psi_\alpha}=\lambda_{\alpha}\ket{\Psi_\alpha}$ can equivalently be written as a zero-variance condition~\cite{Chertkov2018},
\begin{equation}\label{eq:variance}
    \braket{\Delta O^2}_{\alpha} \equiv \braket{\Psi_\alpha|O^2|\Psi_\alpha} - \braket{\Psi_\alpha|O|\Psi_\alpha}^2 = 0 .
\end{equation}
Substituting $O=\sum_i c_i W_i$ yields the quadratic constraint $\mathbf{c}^T M^\alpha \mathbf{c}=0$, where
$M_{ij}^{\alpha}=\braket{\Psi_\alpha|W_iW_j|\Psi_\alpha}-\braket{\Psi_\alpha|W_i|\Psi_\alpha}\braket{\Psi_\alpha|W_j|\Psi_\alpha}$ is the Hermitian operator covariance matrix for state $\ket{\Psi_\alpha}$.
The \textit{parent operator subspace} $\mathcal{K}$ is therefore defined as the joint kernel of covariance matrices $\{M^\alpha\}$~\cite{Chertkov2018, Garrison2018, Qi2019, Rattacaso2024, Bairey2019},
\begin{equation}
    \mathcal{K}=\{\, O=\sum_i c_i W_i \mid \forall \alpha, M^{\alpha}\mathbf{c}=0 \,\}. \label{eq:kernel}
\end{equation}
As quantified in Fig.~\ref{fig2}, $\mathcal{K}$ is typically highly degenerate ($D_\mathcal{K}\gg 1$). While it contains the local parent Hamiltonian and symmetry operators, generic linear combinations within $\mathcal{K}$ are dense mixtures that obscure locality and lack direct physical interpretability.

\subsection{Operator Selection via Parsimony}
The high-dimensional degeneracy of $\mathcal{K}$ necessitates a selection principle to disentangle physically meaningful operators from their non-local mixtures.
Guided by parsimony, we seek the sparsest basis of $\mathcal{K}$.
Since our universal operator basis $\mathcal{W}$ consists of few-body Pauli strings, spatial locality translates into sparsity of the coefficient representation.
Let $K_0\in\mathbb{R}^{N_{\mathcal{W}}\times D_{\mathcal{K}}}$ denote an arbitrary initial basis of $\mathcal{K}$ in this coordinate system.
We then seek an invertible change of basis $R\in \mathrm{GL}(D_{\mathcal{K}})$ such that $K=K_0R$ has minimal support:
\begin{equation}
\begin{aligned}
\min_{R\in \mathrm{GL}(D_{\mathcal{K}})}\quad & \sum_{j=1}^{D_{\mathcal{K}}}\|(K_0R)_j\|_0 \\
\text{s.t.}\quad & \forall j,\ \|(K_0R)_j\|_2=1.
\end{aligned}
\label{eq:optimization}
\end{equation}
where $(\cdot)_j$ denotes the $j$-th column of a matrix, $\|\cdot\|_0$ counts nonzero entries, and the column normalization fixes the overall scale (immaterial for $\ell_0$ but essential for the continuous relaxations below).
This formulation preserves completeness while isolating individual physical operators as maximally sparse columns.

Minimizing the $\ell_0$ objective is NP-hard. Drawing on compressed sensing~\cite{Candes2006, Donoho2006} and dictionary-learning techniques~\cite{Zhang2015}, we solve Eq.~\eqref{eq:optimization} approximately via a robust relaxation strategy (Appendix~\ref{appendix:details}), thereby disentangling $\mathcal{K}$ into local physical generators.
A representative sparse operator dictionary produced by this procedure is shown in Table~\ref{tab:symmetry}.

\subsection{Entropy Selection Criterion}
Sparsity isolates local operators but does not, by itself, single out the Hamiltonian from other conserved quantities.
We therefore invoke a maximal spectral-resolution criterion.
Symmetry generators typically partition the Hilbert space into sectors and hence display spectral degeneracy, whereas the Hamiltonian encodes system-specific interaction connectivity and tends to yield a more resolved spectrum.
We quantify spectral complexity by the Shannon entropy
$S(O)=-\sum_k p_k \ln p_k$,
where $p_k=d_k/\sum_j d_j$ and $d_k$ is the multiplicity of the $k$-th distinct eigenvalue of $O$ within the subspace spanned by the sampled low-lying eigenstates.
We identify the Hamiltonian as the element in the optimized sparse basis of $\mathcal{K}$ that maximizes $S(O)$ (Table~\ref{tab:entropy}).

\begin{table}[b]
\caption{Spectral entropy computed from $5$ low-lying eigenstates. Here $O=\bm{\sigma}_{i}\cdot\bm{\sigma}_{j}$ acts on a geometrically symmetric pair $(i,j)$, i.e., exchanging $i$ and $j$ leaves the interaction graph invariant.
The low-energy states split into bond singlet ($-3$) and triplet ($+1$) sectors,
giving low spectral entropy for $O$,
while $H$ is more spectrally resolved and thus has higher entropy.}

\label{tab:entropy}
\begin{ruledtabular}
\begin{tabular}{lcc}
Operator & $O=\bm{\sigma}_{i}\cdot\bm{\sigma}_{j}$ & $H$ \\
\hline
Eigenvalues & \begin{tabular}{@{}c@{}} $1.0000$ \\ $1.0000$ \\ $-3.0000$ \\ $-3.0000$ \\ $-3.0000$ \end{tabular} & \begin{tabular}{@{}c@{}} $-24.4701$ \\ $-23.5853$ \\ $-22.4165$ \\ $-20.6129$ \\ $-20.4194$ \end{tabular} \\
\hline
Entropy     & $0.6730$ & $1.6094$ 
\end{tabular}
\end{ruledtabular}
\end{table}

\section{Numerical Results}
\subsection{Geometry Reconstruction and Symmetry Discovery} We validate the framework using a geometry-agnostic protocol on the Erd\H{o}s--R\'enyi Heisenberg model (Eq.~\eqref{eq:hamiltonian}).
Treating the interaction geometry as unknown, we task the algorithm with rediscovering the connectivity solely from the operator covariance matrices $\{M^\alpha\}$ computed from low-lying eigenstates.
We first identify the parent operator subspace $\mathcal{K}$ via Eq.~\eqref{eq:kernel}, then disentangle it by sparse basis optimization according to Eq.~\eqref{eq:optimization} and select the Hamiltonian by maximizing spectral entropy.
The interaction geometry is then read off from the support of the recovered Hamiltonian.

\begin{table}[h]
\caption{\label{tab:symmetry} Typical sparse parent operator basis discovered by the optimization algorithm. Rows are colored by class: \colorbox{blue!30}{Hamiltonian}, \colorbox{gray!50}{Intrinsic Symmetries and Redundancies}, and \colorbox{red!30}{Geometric Symmetries}.
Note that $i, j, k \in \{1, \ldots, N_v\}$ range over all sites,
while $m$ and $n$ denote a pair of vertices related by a graph automorphism that swaps $m$ and $n$ while leaving the interaction graph invariant.
}
\begin{ruledtabular}
\begin{tabular*}{.99\linewidth}{l@{\extracolsep{\fill}}l@{\extracolsep{\fill}}l}
       & symmetry operator                              & number                   \\
\hline
\rowcolor{blue!30}
$H$  & $\sum_{\langle i,j\rangle\in G}J_{ij}\boldsymbol{\sigma}_i\cdot\boldsymbol{\sigma}_j$                                            & 1                        \\
\rowcolor{gray!50}
$A$  & $\sum_i \sigma_i^z$                                 & 1                        \\
\rowcolor{gray!50}
$B_i$  & $\sigma_i^z\sum_{j\neq i}\sigma_j^z$                     & $N_{v}$               \\
\rowcolor{gray!50}
$C_{ij}$  & $\sigma_i^z+\sigma_j^z+\sigma_i^z\sigma_j^z\sum_{k\neq i,j}\sigma_k^z$  & $N_{v}(N_{v}-1)/2$ \\
\rowcolor{gray!50}
$D_{ij}$  & $(\sigma_i^x\sigma_j^x+\sigma_i^y\sigma_j^y)\sum_{k\neq i,j}\sigma_k^z$ & $N_{v}(N_{v}-1)/2$ \\
\rowcolor{gray!50}
$E$  & $\sum_{i\neq j}(\sigma_i^x\sigma_j^x+\sigma_i^y\sigma_j^y)$          & 1                        \\
            & total                                     & $(N_{v})^2+3$         \\
\Rowcolor{red!30}
$F$ & $\boldsymbol{\sigma}_m \cdot \boldsymbol{\sigma}_n$                & 1                        \\
\Rowcolor{red!30}
$G$ & $(\boldsymbol{\sigma}_m \cdot \boldsymbol{\sigma}_n - 1)\sigma_k^z, k\neq m,n$    & $N_{v}-2$             \\
\end{tabular*}
\end{ruledtabular}
\end{table}

The sparse parent operator basis yields a physically interpretable operator dictionary, summarized in Table~\ref{tab:symmetry}.
The recovered generators separate into three classes: 
(i) the parent Hamiltonian $H$;
(ii) \emph{intrinsic symmetries and redundancies} (Classes A--E), which are geometry-independent and therefore shared by all interaction graphs in our model class; and
(iii) \emph{geometric symmetries} (Classes F--G), which arise only for special interaction graphs.

Importantly, $H$ and the intrinsic family A--E constitute a geometry-independent baseline shared by all interaction graphs in our model class.
By contrast, the geometric generators appear only when the interaction graph admits nontrivial local automorphisms (i.e., specific symmetric substructures).
These geometry-dependent symmetries account for the excess kernel dimensions beyond the baseline in Fig.~\ref{fig2}, with the corresponding symmetric substructures marked by the coral nodes.

To interpret the discrete plateaus in the kernel dimension $D_{\mathcal K}$ (Fig.~\ref{fig2}), we apply an algebraic generator-extraction step to the sparse basis, compressing the degenerate kernel into a minimal set of fundamental generators. Empirically, $D_{\mathcal K}$ follows a simple structure: a universal baseline $N_v^2+3$ arising from the Hamiltonian and the aforementioned global symmetries and intrinsic spin algebra (Classes A--E), plus a geometry-dependent excess that appears only when the interaction graph admits nontrivial automorphisms (Classes F,G). A concrete factorization for the $D_{\mathcal K}=205$ plateau is given in Appendix~\ref{appendix:details}, where we demonstrate that this high-dimensional parent operator subspace collapses to just eight fundamental generators, most notably the parent Hamiltonian and a hidden long-range exchange operator.

Geometric generators emerge strictly when the interaction graph possesses discrete automorphisms,
causing the kernel dimension to deviate from the intrinsic baseline.
Specifically, the graph corresponding to the $D_{\mathcal{K}} = 200$ plateau (Fig.~\ref{fig2}, left) is invariant under the vertex permutation of $m$ and $n$. This local $\mathbb{Z}_2$ symmetry yields the exchange operator $F=\bm{\sigma}_m \cdot \bm{\sigma}_n$ as a permanent conserved quantity. Furthermore, constraints on the low-energy manifold can induce conditional generators. If the sampled eigenstates are confined to the symmetric triplet sector of the $(m,n)$ bond, the operator $G=(\bm{\sigma}_m \cdot \bm{\sigma}_n - 1)\sigma_i^z$ vanishes identically when acting on this subspace. Consequently, $G$ acts as a null operator with zero variance, thereby entering the kernel as a distinct, state-dependent generator.

\begin{figure}[t]
    \centering
    \includegraphics[width=.97\linewidth]{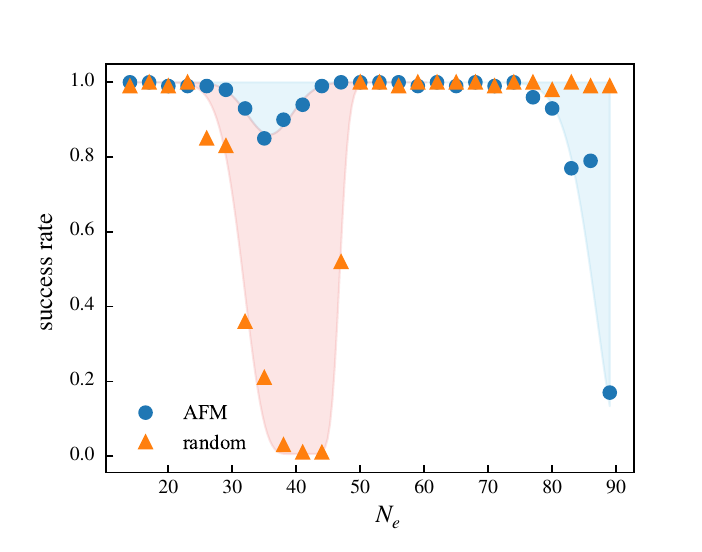}
    \caption{\label{fig3}
    \textbf{Phase diagram of Hamiltonian learnability.}
The success rate of reconstruction is plotted as a function of the number of edges $N_e$.
We compare two interaction classes: the antiferromagnetic (AFM) model (blue circles), where all non-zero coupling constants are fixed at $J_{ij}=1$; and the random-sign model (orange triangles), where interactions $J_{ij}$ are chosen from $\{+1, -1\}$ with equal probability.
The dip in the intermediate region indicates the regime where geometric duality confuses the sparsity criterion.
}
\end{figure}

\subsection{Learnability Phase Diagram} We systematically assess how robustly the interaction geometry can be inferred as a function of edge density of the interaction graph.
The reconstruction success rate in Fig.~\ref{fig3} thus defines a learnability phase diagram for physical laws.
Throughout, we select the parent Hamiltonian using the spectral-entropy maximization criterion introduced in the previous section.

Notably, the spectral-entropy criterion can admit closely competing candidates: multiple sparse operators in $\mathcal{K}$ may achieve indistinguishable entropy on the sampled low-lying states.
In this regime, the data may only constrain the Hamiltonian up to shifts by local conserved operators in $\mathcal{K}$.
For example, if a local conserved operator $F$ is present, the shifted operator $H' = H - \alpha F$ can be indistinguishable from $H$ on the sampled low-lying states while inducing the same interaction geometry, and it may even appear sparser when the $F$-terms cancel a subset of couplings of $H$.
Accordingly, we evaluate performance at the level of geometry recovery: a reconstruction is deemed successful if we identify a parsimonious representative whose spectral entropy is at least that of the ground-truth Hamiltonian and which faithfully captures the essential interaction geometry.

The learning curve exhibits three distinct regimes, with a pronounced ``confusion dip'' in the intermediate-density region ($N_e \approx 40$).
We attribute this anomaly to a competing sparse description supported primarily on the complement graph.
By linearly combining intrinsic symmetries (specifically $E - \sum_i B_i$), one can form an alternative sparse operator
$H' = \sum_{i\neq j} (\sigma_i^x \sigma_j^x + \sigma_i^y \sigma_j^y) - H$.
Unlike the physical Hamiltonian $H$, which is supported on present bonds, this dual candidate is dominated by interactions on the missing bonds.
For random interactions, an analytical comparison of sparsity reveals a parsimony crossover~\footnote{
We define the normalized complexity ratio $\xi = \mathbb{E}[\Vert O \Vert_1 / \Vert O \Vert_2]$.
For the random-sign Hamiltonian, $\xi_H \approx 3N_e/\sqrt{5N_e}$.
Conversely, for the dual candidate $H'$, which exploits the sparsity of the graph complement, the complexity follows $\xi_{H'} \approx (2P+N_e)/\sqrt{4P+5N_e}$, where $P = \binom{N_v}{2}$ is the total number of possible edges.
Solving $\xi_H = \xi_{H'}$ yields a critical density $N_e \approx 0.535 P$ ($N_e \approx 48.7$ for $N_v=14$).
}.
Near this threshold, describing the system in terms of its vacancies becomes mathematically more economical than describing it in terms of its bonds.
Consequently, Occam's Razor favors this dual coordinate system, leading to the observed misidentification near the theoretical intersection.

The limiting behaviors further clarify the physics of reconstruction.
In the \textit{sparse limit} ($N_e < 30$), locality is sharply defined and the Hamiltonian remains the uniquely sparse generator, enabling near-perfect recovery.
In the \textit{dense limit} ($N_e \approx 80$), the system approaches mean-field all-to-all connectivity; correspondingly, the notion of ``local geometry'' loses operational meaning, and the kernel becomes highly degenerate.
Finally, for the random-sign model (orange triangles), the overall success rate is lower than in the antiferromagnetic case.
This suggests that sign disorder reduces the identifiability of the underlying geometry, while the detailed mechanism---and its connection to the sparsity crossover discussed above---merits further study.

\section{Conclusion}
O-Sensing shows that a local Hamiltonian and a set of physically meaningful conserved operators, defined through the sampled states, can be recovered from only a few low-lying eigenstates even when the interaction graph is not provided. The key step is to use parsimony as a gauge-fixing principle within the degenerate parent-operator subspace. By selecting a maximally sparse operator basis, the protocol isolates interpretable generators, and the interaction geometry is identified from the support of the recovered Hamiltonian in the underlying few-body operator expansion. The same sparsity principle also explains the learnability phase diagram. At intermediate graph densities, a comparably sparse dual description supported on the complement graph can become preferred, leading to the observed ``confusion regime'' in which parsimony selects the dual representation.

From an experimental standpoint, the main requirement is access to the few-body correlators entering the operator covariance matrices. Recent progress in randomized measurements, especially classical-shadow tomography, offers a practical way to estimate many such correlators with sample complexity that grows only logarithmically with the number of target observables~\cite{Huang2020,Hu2023,Ippoliti2023}.
While these techniques are promising, the protocol's performance under experimental noise and finite-sampling errors still needs to be quantified. Quantifying the noise tolerance of the reconstructed geometry is therefore a critical next step for implementation on near-term quantum hardware.

\appendix
\section{Algorithm Details}\label{appendix:details}

\subsection{Mathematical Formalism of Sparse Optimization}
We briefly set up the mathematical formulation underlying the sparse optimization.
We work in the geometry-independent operator basis $\mathcal{W}$ defined in the main text.
Given sampled stationary states $\{\ket{\Psi_\alpha}\}$, we define the operator variance $\mathcal{V}_{\alpha}(O)=\braket{\Psi_\alpha|O^2|\Psi_\alpha}-\braket{\Psi_\alpha|O|\Psi_\alpha}^2$ and the corresponding parent operator subspace $\mathcal{K}^{\alpha}=\{\,O\in\mathcal{W}\mid \mathcal{V}_{\alpha}(O)=0\,\}$.
When multiple stationary states $\{\ket{\Psi_\alpha}\}$ are used, we take $\mathcal{K}$ to be the joint kernel, i.e., $\mathcal{V}_{\alpha}(O)=0$ for all $\ket{\Psi_\alpha}$.

As a working viewpoint, it is helpful to regard the subspace $\mathcal{K}$ as the Hermitian sector of the intersection of an underlying algebraic structure $\mathcal{A}$ associated with the system's symmetries.
Concretely, let $\mathcal{G}=\{G_1,G_2,\ldots\}$ denote a minimal set of physical generators (e.g., the Hamiltonian and irreducible symmetry operators), and take $\mathcal{A}$ to be the unital associative algebra generated by $\mathcal{G}$.
In practice, this means that $\mathcal{A}$ contains the identity and all finite products of the generators and their linear combinations (e.g., $\mathrm{span}\{\mathbb{I}, G_1, G_2, G_1G_2,\ldots\}$).
Since physical observables are necessarily Hermitian, we restrict attention to the Hermitian sector of $\mathcal{A}$---generic products such as $G_1G_2$ need not be Hermitian. Instead, we heuristically interpret the observed subspace $\mathcal{K}$ as the Hermitian slice:
\begin{equation}
\mathcal{K} = \left\{ O \in \mathcal{A} \cap \mathcal{W} \;\middle|\; O = O^\dagger \right\}.
\end{equation}
Within this framework, extracting $\mathcal{G}$ from $\mathcal{K}$ is generally non-unique due to basis ambiguity: any linear combination of conserved quantities remains conserved.
We adopt sparsity as a practical selection principle.
The guiding intuition is that, although $\mathcal{K}$ contains many non-local algebraic composites, physically meaningful generators tend to admit comparatively sparse representations (maximal locality) in the chosen basis.
Accordingly, the algorithm searches for basis rotations of $\mathcal{K}$ that promote sparsity (approximating $\ell_0$-minimization), thereby suppressing algebraic redundancies.

To rigorously define sparsity, we transform the abstract operators to coefficient vectors.
With the Hilbert--Schmidt inner product $\langle A,B\rangle \equiv \mathrm{Tr}(A^\dagger B)$, the geometry-independent basis $\{W_i\}_{i=1}^{N_{\mathcal W}}$ is chosen to be orthonormal, $\mathrm{Tr}(W_i^\dagger W_j)=\delta_{ij}$.
Any candidate operator $O$ is then uniquely specified by its coefficient vector $\mathbf{k}\in\mathbb{R}^{N_{\mathcal W}}$:
\begin{equation}
O=\sum_{i=1}^{N_{\mathcal W}} k_i W_i, 
\qquad 
k_i=\langle W_i,O\rangle=\mathrm{Tr}(W_i^\dagger O).
\end{equation}
We denote the dimension of this subspace $\mathcal{K}$ by $D_\mathcal{K}$. Numerically, we obtain an initial orthonormal basis $\mathcal{B}_0 = \{ \mathbf{k}^{(0)}_1, \dots, \mathbf{k}^{(0)}_{D_\mathcal{K}} \}$ spanning this subspace via the null-space of the covariance matrix. We organize these basis vectors as columns of a matrix $K_0 \in \mathbb{R}^{N_\mathcal{W} \times D_\mathcal{K}}$. Since the physical operators---the Hamiltonian and symmetries---essentially constitute a specific, sparse basis of $\mathcal{K}$, our task effectively becomes a basis rotation problem.

We seek a linear transformation matrix $R \in \text{GL}(D_\mathcal{K}, \mathbb{R})$ such that the transformed basis $K = K_0 R$ contains as many zero entries as possible. The columns of $K$ represent the recovered physical candidates. The optimization problem is thus formulated as finding the optimal $R$ that minimizes the $\ell_0$-norm of the columns of $K$:
\begin{equation}
    \min_{R \in \text{GL}(D_\mathcal{K})} \sum_{j=1}^{D_\mathcal{K}} \| (K_0 R)_j \|_0, \quad \text{s.t.} \quad \| (K_0 R)_j \|_2 = 1,
    \label{eq:sparse_obj}
\end{equation}
where $(\cdot)_j$ denotes the $j$-th column vector. The normalization constraint removes the trivial scaling ambiguity. Since physical conservation laws are invariant under sign inversion---$O$ and $-O$ are physically equivalent---the solution is unique only up to a signed permutation. Consequently, we consider the physical laws successfully recovered if the identified basis $K$ equals $K^* P \Sigma$, where $K^*$ is the ground truth, $P$ is a permutation matrix, and $\Sigma$ is a diagonal sign matrix.

The optimization objective \eqref{eq:sparse_obj} employs the $\ell_0$-pseudonorm to count the number of non-zero terms in the operator dictionary. However, due to its discrete nature, we employ continuous proxies based on the $\ell_p$-norm, defined as $\|\mathbf{k}\|_p = (\sum_{i} |k_i|^p)^{1/p}$. In practice, we exploit two distinct geometric properties: maximizing the $\ell_3$-norm ($p>2$) favors ``spiky'' distributions on the sphere to achieve rapid approximate identification, while minimizing the $\ell_1$-norm ($p=1$) serves as a robust, continuous surrogate for sparsity, refining the solution by suppressing small coefficients, analogous to the relaxation used in compressed sensing.

\begin{figure}[b]
\centering
    \centering
    \begin{overpic}[width=0.98\linewidth]{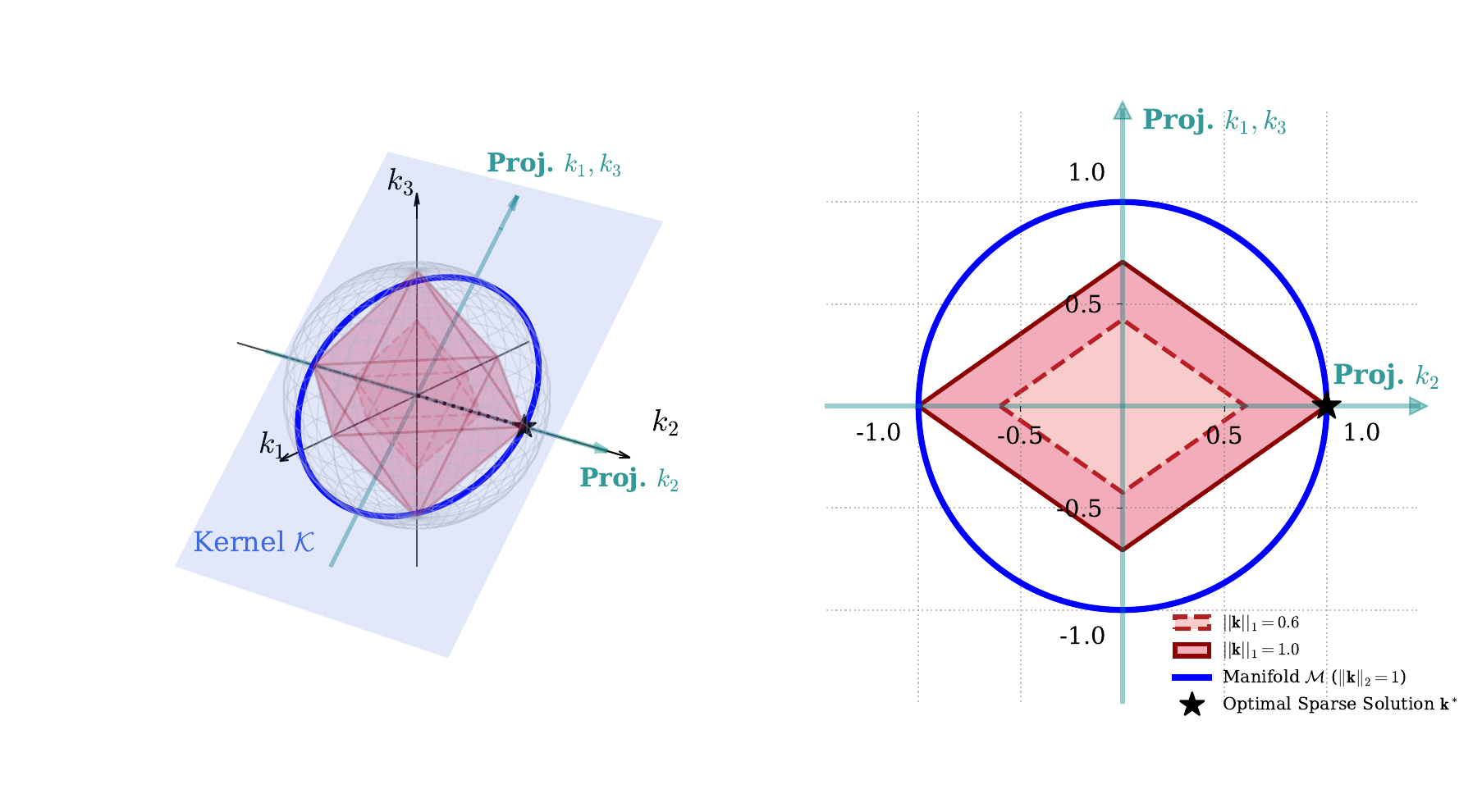}
        \put( 10, 105){\textbf{(a)}}
        \put(120, 105){\textbf{(b)}}
    \end{overpic}
\caption{\textbf{Geometric mechanism of sparse operator discovery.} 
\textbf{(a)} Ambient space view: The valid solution manifold $\mathcal{M}$ (blue ring) is determined by the intersection of the physical kernel subspace $\mathcal{K}$ (semi-transparent blue plane) and the $\ell_2$-normalization sphere (light gray). The teal arrows define the intrinsic projection coordinates within $\mathcal{K}$. 
\textbf{(b)} Subspace projection: Sparse optimization visualized within the $\mathcal{K}$-plane. The $\ell_1$-norm level sets form anisotropic cross-polytopes (red diamonds). Minimizing the $\ell_1$-norm is geometrically equivalent to identifying the minimal $\ell_1$-polytope that makes contact with $\mathcal{M}$. Because of the sharp vertices of the $\ell_1$-ball, this contact strictly occurs along a coordinate axis, naturally collapsing redundant degrees of freedom to isolate the optimal sparse solution $\mathbf{k}^*$ (black star).}
\label{fig:sparse_opt}
\end{figure}

\subsection{Geometric Interpretation of Sparse Optimization}
To intuitively understand why minimizing the $\ell_1$-norm successfully recovers a local, sparse basis from the highly degenerate kernel space, we consider the geometric landscape illustrated in Fig.~\ref{fig:sparse_opt}.

The problem is defined in a high-dimensional Cartesian system spanned by the geometry-independent basis $\{W_i\}$ [Fig.~\ref{fig:sparse_opt}(a)]. Within this ambient space, the physical constraints impose two geometric conditions: the stationarity condition ($\langle \Delta O^2 \rangle = 0$) confines valid operators to a linear subspace (the Kernel $\mathcal{K}$), while the normalization constraint ($\|\mathbf{k}\|_2=1$) restricts them to the unit sphere. Consequently, any valid parent operator must reside on the intersection of these two structures, forming the solution manifold $\mathcal{M}$ (shown as the blue ring).

To single out physically interpretable operators from this manifold, we minimize the $\ell_1$-norm. As shown in the projected view of Fig.~\ref{fig:sparse_opt}(b), the unit ball of the $\ell_1$-norm forms a cross-polytope with sharp vertices aligned with the coordinate axes. This spiky geometry makes the $\ell_1$-isosurface tend to intersect the solution manifold $\mathcal{M}$ near axis-aligned vertices. Consequently, the optimal solution $\mathbf{k}^*$ statistically aligns with the canonical axes, forcing off-axis components to vanish and naturally recovering a sparse physical law.

While Fig.~\ref{fig:sparse_opt} illustrates this concept in 3-dimensions, the effect becomes significantly more pronounced in high-dimensional spaces. Strictly speaking, our operator discovery falls into the class of Complete Dictionary Learning problems with a spherical constraint. Recent theoretical advances~\cite{Sun2017, Shen2020} have extended the landscape analysis of sparse optimization to this manifold setting.
These works show that, in complete dictionary learning with spherical constraints, maximizing an $\ell_p$-type spikiness objective (typically with $p > 2$) over the orthogonal group promotes sparsity and can recover the underlying sparse basis~\cite{Shen2020}.
This provides theoretical motivation for our Stage I, where we use $\ell_3$-maximization as a smooth surrogate to rapidly steer the basis toward sparse, interpretable generators.

\subsection{Two-Stage Relaxation Strategy}
Navigating the high-dimensional landscape of sparse optimization is non-trivial; the $\ell_0$-norm is discrete and combinatorial, while the convex $\ell_1$-relaxation, despite its theoretical guarantees, can still suffer from slow convergence or numerical instabilities in specific manifold settings. To overcome these barriers, we engineer a two-stage protocol that leverages distinct geometric properties of norm landscapes to balance global exploration with local precision.

\textbf{Stage I: Global Exploration.}
We initiate the search by maximizing the $\ell_3$-norm. Geometrically, on the $\ell_2$ unit sphere, maximizing $\|\cdot\|_p$ with $p>2$ produces an axis-peaked objective landscape whose maxima occur at coordinate-aligned sparse directions.
We initiate Stage I by solving the orthogonally constrained spikiness maximization
\begin{equation}
\max_{A\in O(D_{\mathcal K})}\ \|K_0A\|_3^3 .
\end{equation}
Starting from an initial $A^{(0)}$, each iteration forms the ascent matrix associated with this $\ell_3$ objective as in Ref.~\cite{Sun2017}, and projects it back onto the orthogonal group by an SVD. Concretely, if the ascent matrix has singular value decomposition $U\Sigma V^\top$, we update
\begin{equation}
A^{(t+1)}=UV^\top ,
\end{equation}
and set $K^{(t+1)}=K_0A^{(t+1)}$. This produces an orthogonal rotation that biases the kernel basis toward sparse, physically interpretable generators.

\textbf{Stage II: Local Refinement.}
Starting from the globally rotated basis obtained in Stage I, we further sparsify the operators. The key step is to relax the mutual orthogonality constraint. By dropping the requirement that different basis vectors remain orthogonal, we decompose the refinement into $D_{\mathcal{K}}$ independent sub-problems. This allows each operator to independently rotate on the unit sphere and search for its own maximally sparse representation.

Mathematically, we construct each candidate operator $\mathbf{k}$ as a linear superposition of the Stage I basis vectors $\{\mathbf{v}_l\}$. For each operator, we optimize the mixing coefficients $\mathbf{q}=(q_1,\dots,q_{D_{\mathcal{K}}})^\top$ by solving
\begin{equation}
\min_{\mathbf{q}} \left\| \sum_{l=1}^{D_{\mathcal{K}}} q_l \mathbf{v}_l \right\|_1
\quad \text{subject to} \quad
\sum_{l=1}^{D_{\mathcal{K}}} |q_l|^2 = 1.
\end{equation}
The constraint $\|\mathbf{q}\|_2=1$ fixes the overall scale of $\mathbf{k}$ and enforces a unit-$\ell_2$ normalization. We solve these independent problems in parallel using Adam updates with explicit re-normalization after each iteration. This formulation makes it explicit that Stage II searches for sparse directions within the subspace spanned by Stage I, thereby further purifying the operator dictionary.

A potential concern with independent optimization is mode collapse, where multiple basis vectors converge to the same sparse direction. In practice, this is mitigated by the strong separation achieved in Stage I, which provides a well-spread initialization across different basins of attraction. While an explicit incoherence penalty could be added to enforce diversity, we find it unnecessary in our settings.

\subsection{Algebraic Generator Extraction}
To extract a minimal set of fundamental generators, we employ an iterative procedure that filters out algebraically redundant operators. Let the set of sparse operators obtained from Stage II be the candidate pool. We maintain two sets during the process: (i) the set of identified generators $\mathcal{G}$, and (ii) a basis set $\mathcal{B}$ that tracks the linearly independent product combinations generated by $\mathcal{G}$. Both are initialized as empty sets. The procedure is as follows:

\begin{enumerate}
    \item \textbf{Complexity Ordering}: We sort the candidate pool by $\ell_1$-norm complexity in ascending order. This ensures that the sparsest (simplest) operators are prioritized for evaluation.

    \item \textbf{Independence Check}: We iteratively extract the candidate $O_{\mathrm{new}}$ with the minimal $\ell_1$-norm from the remaining pool. We then check if $O_{\mathrm{new}}$ is linearly independent of the current algebraic basis $\mathcal{B}$. That is, we test if $O_{\mathrm{new}}$ can be represented as a linear combination of existing elements in $\mathcal{B}$. If it is linearly dependent, $O_{\mathrm{new}}$ is discarded.

    \item \textbf{Algebraic Expansion}: If $O_{\mathrm{new}}$ is linearly independent, it is identified as a new fundamental generator. We add $O_{\mathrm{new}}$ to the generator set $\mathcal{G}$. Crucially, to maintain the algebraic consistency of the check, we must also expand $\mathcal{B}$ to include the products of the new generator with existing terms.
    For example, if the current basis tracks the algebra of $\{G_1\}$ (i.e., $\mathcal{B}=\{G_1\}$), and we accept a new generator $G_2$, we expand $\mathcal{B}$ to include not just $G_2$, but also its independent cross-products (e.g., updating $\mathcal{B}$ to $\{G_1, G_2, G_1 G_2, G_2 G_1\}$). This ensures that subsequent composite candidates are correctly identified as redundant and rejected.
\end{enumerate}

\begin{figure}[t]
    \centering
    \includegraphics[width=.65\linewidth]{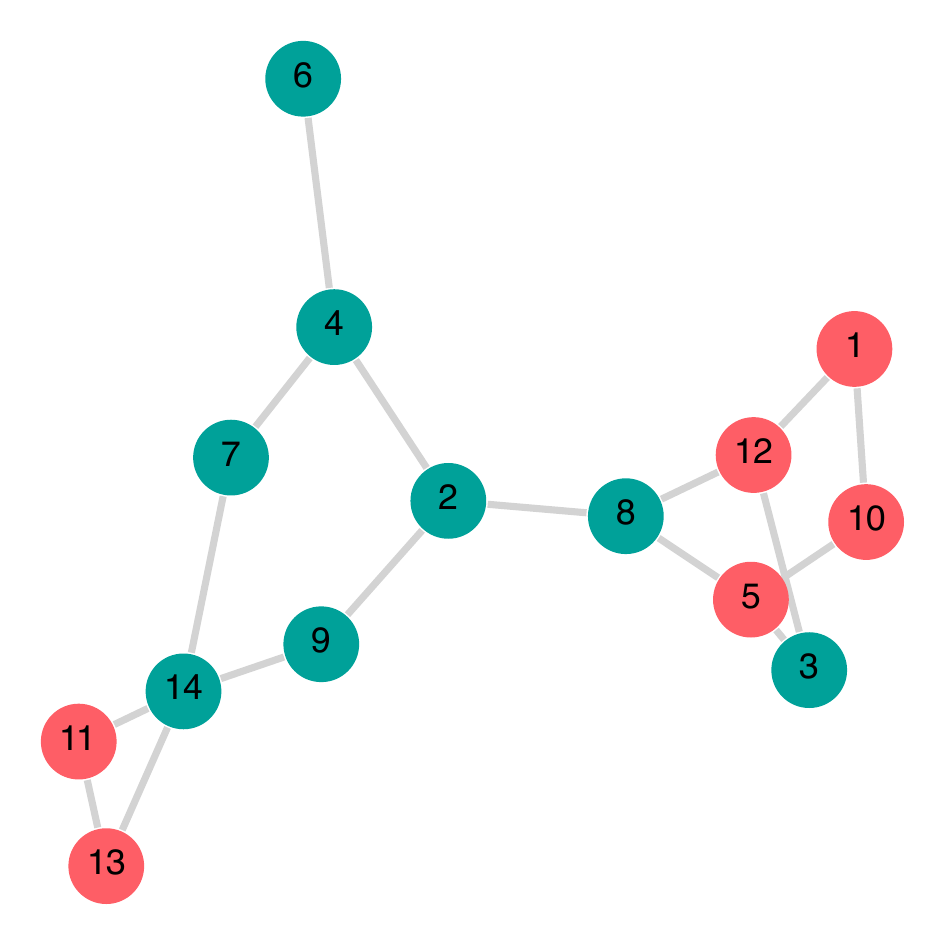}
    \caption{\label{fig:geo205}
    Geometry of the selected instance with $D_{\mathcal{K}}=205$.
}
\end{figure}

To illustrate the practical effect of the above algorithm, we focus on a representative instance on the $D_{\mathcal K}=205$ plateau of Fig.~\ref{fig2}. The corresponding interaction graph, shown in Fig.~\ref{fig:geo205}, exhibits a nontrivial automorphism, providing a concrete geometric setting in which extra conserved structures emerge. Although the raw kernel basis is highly degenerate, applying the algebraic extraction step compresses the parent subspace to eight fundamental generators, listed in Table~\ref{tab:gen205}. In particular, the extracted set isolates the learned Hamiltonian, identifies the geometry-dependent exchange operator associated with the symmetric vertices, and organizes the remaining elements into transparent factorized composites built from these primitives.

\begin{table}[t]
\caption{Algebraically extracted generators for the 205-dimensional parent subspace.
We introduce the shorthand
$X_{ij}\equiv \sigma_i^x\sigma_j^x+\sigma_i^y\sigma_j^y$, where $H$ is the learned Hamiltonian. 
The eight generators fall into a few structural groups.
All coefficients follow the normalization convention of the extracted basis.}
\label{tab:gen205}
\begin{ruledtabular}
\begin{tabular*}{.99\linewidth}{l@{\extracolsep{\fill}}l}
Generator & Compact factorized form \\
\hline
$G_1$ & $\bm{\sigma}_{11} \cdot \bm{\sigma}_{13}$ \\
$G_2$ & $\Bigl(\sigma_1^z+\sigma_{10}^z\Bigr)\,\Bigl(\sigma_5^z+\sigma_{12}^z\Bigr)\,\Bigl(\sigma_{11}^z+\sigma_{13}^z\Bigr)$ \\
$G_3$ & $\Bigl(\sigma_1^z+\sigma_{10}^z\Bigr)\;\Bigl(X_{5,11}+X_{5,13}+X_{11,12}+X_{12,13}\Bigr)$ \\
$G_4$ & $\Bigl(\sigma_5^z+\sigma_{12}^z\Bigr)\;\Bigl(X_{1,11}+X_{1,13}+X_{10,11}+X_{10,13}\Bigr)$ \\
$G_5$ & $\Bigl(\sigma_{11}^z+\sigma_{13}^z\Bigr)\;\Bigl(X_{1,5}+X_{1,12}+X_{10,5}+X_{10,12}\Bigr)$ \\
$G_6$ & $\Bigl(\sigma_{11}^z+\sigma_{13}^z\Bigr)\;\Bigl(\mathbb I+\frac12\,\bm{\sigma}_{1}\cdot\bm{\sigma}_{10}
+\frac12\,\bm{\sigma}_{5}\cdot\bm{\sigma}_{12}\Bigr)$ \\
$G_7$ & $H=\sum_{\langle i,j\rangle\in G}J_{ij}\,\bm{\sigma}_i\cdot\bm{\sigma}_j$ \\
$G_8$ & $\sum_{i\neq j}X_{ij}-4\sqrt{5}\;\Bigl( H-\bm{\sigma}_{11} \cdot \bm{\sigma}_{13} \Bigr)$ \\
\end{tabular*}
\end{ruledtabular}
\end{table}

\section{Detailed Analysis of Table~\ref{tab:symmetry}}

This section elucidates the algebraic origin of the conserved quantities listed in Table~\ref{tab:symmetry}.
The identified operators satisfy the stationarity condition on the sampled low-lying eigenstates $\ket{\Psi_\alpha}$.

\textit{Intrinsic Symmetries (Classes A--E).}
These operators arise from the intrinsic spin algebra and the conservation of total magnetization.

Class A corresponds to the total magnetization operator $A = \sum_k \sigma_k^z$.
In the zero-magnetization subspace, the states satisfy the constraint $A \ket{\Psi} = 0$.
This implies the identity $\sum_{k\neq i,j...} \sigma_k^z \ket{\Psi} = -(\sigma_i^z + \sigma_j^z + \dots) \ket{\Psi}$.

Class B operators reduce to scalars via the Class A constraint:
\begin{align}
    B_i \ket{\Psi}
    &= \sigma_i^z \sum_{j\neq i} \sigma_j^z \ket{\Psi} \nonumber \\
    &= \sigma_i^z (-\sigma_i^z) \ket{\Psi} \nonumber \\
    &= -\mathbb{I} \ket{\Psi}.
\end{align}

Class C operators vanish algebraically upon substitution:
\begin{align}
    C_{ij} \ket{\Psi}
    &= \left[ \sigma_i^z + \sigma_j^z + \sigma_i^z \sigma_j^z \sum_{k\neq i,j} \sigma_k^z \right] \ket{\Psi} \nonumber \\
    &= \left[ \sigma_i^z + \sigma_j^z - \sigma_i^z \sigma_j^z (\sigma_i^z + \sigma_j^z) \right] \ket{\Psi} \nonumber \\
    &= \left[ \sigma_i^z + \sigma_j^z - \sigma_j^z - \sigma_i^z \right] \ket{\Psi} \nonumber \\
    &= 0.
\end{align}

Class D operators vanish identically due to the orthogonality of the spin basis states.
Decomposing the interaction into exchange and polarization terms:
\begin{align}
    D_{ij} \ket{\Psi}
    &\propto (\sigma_i^+ \sigma_j^- + \sigma_i^- \sigma_j^+) (\sigma_i^z + \sigma_j^z) \ket{\Psi} \nonumber \\
    &= 0.
\end{align}
The product vanishes because the polarization factor $(\sigma_i^z + \sigma_j^z)$ annihilates states with antiparallel spins ($\ket{\uparrow\downarrow}, \ket{\downarrow\uparrow}$), while the exchange factor $(\sigma_i^+ \sigma_j^- + \mathrm{h.c.})$ annihilates states with parallel spins ($\ket{\uparrow\uparrow}, \ket{\downarrow\downarrow}$).

Class E derives from the conservation of the total spin squared, $\bm{\sigma}_{\text{tot}}^2 = (\sum_i \bm{\sigma}_i)^2$.
Expanding this operator in terms of Pauli components yields:
\begin{equation}
    \bm{\sigma}_{\text{tot}}^2 = \sum_{i,j} (\sigma_i^x \sigma_j^x + \sigma_i^y \sigma_j^y + \sigma_i^z \sigma_j^z).
\end{equation}
Isolating the transverse terms (involving $x$ and $y$ components) allows us to express $E$ as:
\begin{align}
    E \ket{\Psi}
    &= \sum_{i\neq j} (\sigma_i^x \sigma_j^x + \sigma_i^y \sigma_j^y) \ket{\Psi} \nonumber \\
    &= \left[ \bm{\sigma}_{\text{tot}}^2 - \left(\sum_k \sigma_k^z\right)^2 - \sum_k (\sigma_k^x)^2 - \sum_k (\sigma_k^y)^2 \right] \ket{\Psi} \nonumber \\
    &= \left[ \bm{\sigma}_{\text{tot}}^2 - 0 - 2N_v \mathbb{I} \right] \ket{\Psi}.
\end{align}
Here, we utilized the zero-magnetization constraint $(\sum_k \sigma_k^z)\ket{\Psi}=0$ and the identity $(\sigma_k^x)^2 = (\sigma_k^y)^2 = \mathbb{I}$.
Since the total spin $\bm{\sigma}_{\text{tot}}^2$ is a conserved quantity, it acts as a scalar constant on any subspace with fixed total spin quantum number $S$.
Consequently, $E$ reduces to a trivial scalar operator within this subspace.

\textit{Geometric Symmetries (Classes F--G).}
These operators emerge from the discrete symmetries of the interaction graph.
If the graph is invariant under the vertex permutation between sites $m$ and $n$, the exchange interaction on the symmetric bond, $F_{mn} = \bm{\sigma}_m \cdot \bm{\sigma}_n$, commutes with the Hamiltonian.

Class G arises when the sampled eigenstates are confined to a specific eigensector of the symmetric bond operator $F_{mn}$ (e.g., the triplet sector with eigenvalue $\lambda$).
For any site $k$ distinct from $m$ and $n$, the operator vanishes on these states:
\begin{align}
    G \ket{\Psi}
    &= \sigma_k^z (\bm{\sigma}_m \cdot \bm{\sigma}_n - \lambda) \ket{\Psi} \nonumber \\
    &= 0,
\end{align}
where we have used the fact that $\sigma_k^z$ commutes with the bond operator on sites $m,n$.

\section{Performance on Regular Lattices}

\begin{figure}[ht!]
    \centering
    \includegraphics[width=0.95\linewidth]{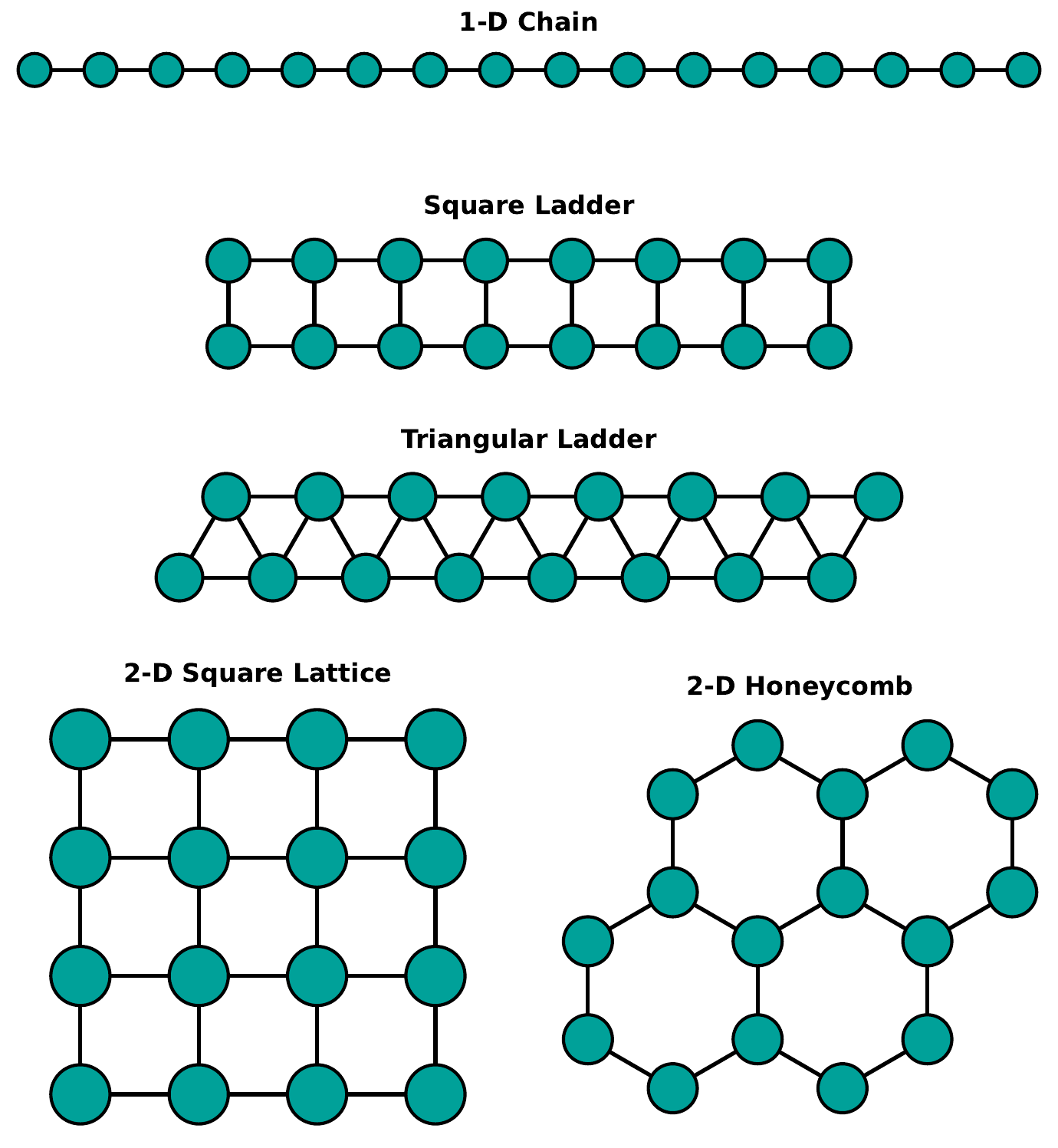}
    \caption{\label{fig:regular}
    \textbf{Schematic of regular interaction geometries.}
    We validate the O-Sensing protocol on diverse regular lattices, including 1D chains, square ladders, triangular ladders, 2D square lattices, and honeycomb lattices.
    In all cases, we take $N_v=16$ and nearest-neighbor antiferromagnetic interactions ($J_{ij}=1$).
    }
\end{figure}

To demonstrate the universality of the O-Sensing protocol, we extended our benchmarks beyond random graphs to standard regular geometries (see Fig.~\ref{fig:regular}).
Specifically, we tested the algorithm on 1D chains, 1D rings, square ladders, triangular ladders, 2D square lattices, and honeycomb lattices.
In all these benchmarks, we fixed the system size to $N_v=16$ sites and employed nearest-neighbor antiferromagnetic interactions ($J_{ij}=1$).

In all tested instances, the algorithm successfully reconstructed the exact parent Hamiltonian and the associated conserved quantities.
The dimension of the recovered parent operator subspace was consistently found to be $N_v^2+3$ (which equals $259$ for $N_v=16$).
This result aligns perfectly with the theoretical expectation for generic Heisenberg models lacking local exchange symmetries: the kernel consists strictly of the Hamiltonian, the global spin algebra (total spin operators), and the intrinsic redundancies (Classes A--E).
The accurate recovery of this baseline dimension across diverse geometries confirms that O-Sensing is geometry-agnostic, reliably identifying the correct physical laws for various lattices without prior knowledge of the underlying structure.

\bibliography{references}

\end{document}